\documentclass[12pt,preprint]{aastex}

\usepackage{epsf}

\newcommand{\tx}[1]{\textrm{#1}}

\newcommand{\kms}{km~$\tx{s}^{-1}$}
\newcommand{\dv}{$r^{1/4}\,$}

\newcommand{\mbh}{M$_{\rm BH}$}

\slugcomment{ApJ Letters, submitted}

\shorttitle{Black hole mass $\sigma$ at $z\sim0.37$ }
\shortauthors{Treu, Malkan \& Blandford}

\begin{document}
\title{The relation between black hole mass and velocity dispersion at $z\sim0.37$
}

\footnotetext[1]{Based on data collected at Keck
Observatory, operated by Caltech and the University of California.}

\author{Tommaso Treu$\!$\altaffilmark{2,3,4}, Matthew A. Malkan $\!$\altaffilmark{2} \& Roger D. Blandford $\!$\altaffilmark{4,5}}
\altaffiltext{2}{University of California at Los Angeles, CA 90095; ttreu@astro.ucla.edu, malkan@astro.ucla.edu}
\altaffiltext{3}{Hubble Fellow} 
\altaffiltext{4}{California Institute of Technology. Pasadena, CA 91125}
\altaffiltext{5}{Kavli Institute for Particle Astrophysics and Cosmology
Stanford, CA; rdb@slac.stanford.edu}

\begin{abstract}
The velocity dispersion of 7 Seyfert 1 galaxies at $z\sim0.37$ is
measured using high signal-to-noise Keck spectra.  Black hole (BH)
mass estimates are obtained via an empirically calibrated
photoionization method. We derive the BH mass velocity dispersion
relationship at $z\sim0.37$. We find an offset with respect to the
local relationship, in the sense of somewhat lower velocity dispersion
at a fixed BH mass at $z\sim0.37$ than today, significant at the 97\%
level. The offset corresponds to $\Delta \log \sigma$ = $-0.16$ with
rms scatter of 0.13 dex. If confirmed by larger samples and
independent checks on systematic uncertainties and selection effects,
this result would be consistent with spheroids evolving faster than
BHs in the past 4 Gyrs and inconsistent with pure luminosity
evolution.
\end{abstract}


\section{Introduction}

The correlation of the mass of the central black hole (M$_{\rm BH}$)
with the spheroid velocity dispersion $\sigma$ (Ferrarese \& Merritt
2000; Gebhardt et al.\ 2000; hereafter BHS) links phenomena at widely
different scales (from the pcs of the BH sphere of influence to the
kpcs of the bulge).  This demonstrates that galaxy formation
and AGN activity are connected and several physical explanations have
been proposed (e.g. Kauffmann \& Haenhelt 2000; Monaco et al.\ 2000;
Volonteri et al.\ 2003; Haiman, Ciotti \& Ostriker 2004).

Interesting clues can be obtained from the cosmic evolution of
empirical relations. Different scenarios -- all reproducing the local
BHS relation -- predict different evolution. For example, in a pure
luminosity evolution scenario of spheroids, $\sigma$ would not change
with time, while M$_{\rm BH}$ would increase as a result of
accretion. A typical $10^7 M_{\odot}$ BH accreting at an average rate
$0.01 M_{\odot} yr^{-1}$ (e.g., Sun \& Malkan 1989), would double its
mass in a Gyr. This would predict the \mbh\, at given $\sigma$ to
increase with time, possibly changing the slope of the BHS relation if
the accretion rate is a function of M$_{\rm BH}$ (e.g. Small \&
Blandford 1992). In contrast, if spheroids grew faster than BH,
$\sigma$ at fixed M$_{\rm BH}$ could increase with time.

In this {\it Letter} we present the first results from an
observational program aimed at measuring the cosmic evolution of the
BHS relation. We measure the BHS relation for a sample of 7 Seyfert 1s
at $z\sim0.37$. This redshift is high enough to provide a time
baseline over which we might expect evolution, yet low enough to be
observationally practical.

\section{Observations and analysis}

\subsection{Experiment Design and Sample Selection}

The sphere of influence of supermassive BHs in galaxies at
cosmological distances cannot be resolved even with the Hubble Space
Telescope (HST). Therefore we target active galaxies, where \mbh\, can
be obtained from the integrated properties of the broad emission-line
region. In this paper, we will combine an empirically calibrated
photo-ionization estimate of the size of the broad line region
(hereafter ECPI; Wandel et al. 1999) with its kinematics measured from
the rms width of H$\beta$, to deduce the central BH mass.  To measure
simultaneously the {\it stellar velocity dispersion} from absorption
lines, requires nuclei of relatively low luminosity, so that the
fraction of stellar light in the integrated spectrum is substantial.
Seyfert 1s provide the right balance between the two components:
absorption features typical of old stellar populations such as Mg5175
and Fe5270 are clearly visible in their high signal-to-noise
integrated spectra. In order to minimize the uncertainties from sky
subtraction and atmospheric absorption corrections, it is convenient
to select specific redshift windows where the relevant emission and
absorption lines fall in clean regions of the atmosphere. Accordingly,
we selected the ``clean window'' $z\sim0.37$, which corresponds to a
look-back time of $\sim 4$ Gyrs, for
H$_0=70$~km\,s$^{-1}$\,Mpc$^{-1}$, $\Omega_{\rm m}=0.3$, and
$\Omega_{\Lambda}=0.7$.

A first object (MS1558+453; hereafter S99; Stocke et al.\ 1991) was
selected for a pilot study. When the Sloan Digital Sky Survey (SDSS)
became available, a larger sample of objects was selected according to
the following criteria: $0.35<z<0.37$, H$\beta$ equivalent width and
rms width greater than 5 \AA. These are sufficient to select only
broad H$\beta$ galaxies, but are loose enough that they should not
introduce significant bias in \mbh. The relevant properties of the
observed objects are listed in Table~\ref{tab:sample}.

\subsection{Observations and data reduction}

High signal-to-noise (Table~\ref{tab:sample}) spectra were obtained
using the Low Resolution Imaging Spectrograph (Oke et al. 1995) at the
Keck-I telescope on 2003 March 6 (S99) and 2003 Sep 3. Two exposures
were obtained for each object, with total exposure times ranging
between 1200s and 5280s. The 900/5500 grating with a $1\farcs5$ slit
provided a resolution of $\sigma_s= 55\pm5$ kms$^{-1}$ around Mg5175
and Fe5270, as measured from sky lines and arc lamps. Internal flat
fields were obtained after each object exposure, to correct the
fringing pattern of the red CCD. A set of A0V stars from the Hipparcos
catalog\footnote{URL
http://www.gemini.edu/sciops/instruments/niri/NIRISpecStdSearch.html}
to be $\lesssim$ 15 degrees from each target was observed during the
night as flux calibrators, and to measure the B-band atmospheric
absorption. Spectrophotometric standards were observed during
twilights. Internal tests, and comparison with SDSS spectra, show that
this procedure corrects the B-band absorption to a level of a few
parts in a thousand and relative flux calibration to a few
percent. The data reduction was similar to that in Treu et al. (2001).

\subsection{Bulge Kinematics}

The stellar velocity dispersion of the bulge was obtained by comparing
the spectral region corresponding to 5100-5300 \AA\ with spectra of
G-K giants as described in Treu et al.\ (2001). Briefly, the high
resolution template stars were redshifted and smoothed to match the
resolution of the instrumental setup, and convolved with gaussians in
$\log \lambda$ space to reproduce the kinematic broadening. Then a low
order polynomial representing the featureless continuum was added
(using the pixel fitting code by van der Marel 1994). Small AGN
emission features at vacuum wavelengths $\lambda$5160.33,
$\lambda$5200.53, and $\lambda$5310.34 \AA\, (van den Berk et
al. 2001) were masked out during the fit. This procedure yields a
velocity dispersion $\sigma_{\rm ap}$, a line strength $\gamma$, with
uncertainties (Tab.~1). For each object, we performed a number of
tests, varying the spectral range and the order of the polynomial used
for continuum fitting. The results were found to be sensitive to these
changes for 6/13 objects (the ones with shallower stellar absorption
features and typically with stronger FeII AGN emission). The other 7
objects (Fig.~2) yielded stable $\sigma$ -- changes much smaller than
estimated errors -- and were therefore considered reliable.  We note
that excluding the Mg region changes the velocity dispersion by less
than the estimated errors (c.f. Barth, Ho \& Sargent 2003). For
simplicity and consistency with previous work we assume that the
central velocity dispersion $\sigma$ (i.e. within a circular aperture
of radius 1/8 of the effective radius; Ferrarese \& Merritt 2000) can
be obtained as $\sigma$ = ${\mathcal B} \sigma_{ap}$, with ${\mathcal
B}$=$1.1\pm0.05$. This assumes that the spheroids have an effective
radius of $0\farcs5$, are non rotating, their velocity dispersion
profile is similar to that of early-type galaxies, and the disk
contamination to the line profile is negligible.

\subsection{Black hole mass determination}

Black hole masses were determined using the empirical correlation
between continuum luminosity and size of the broad line region (Wandel
et al. 1999; Kaspi et al. 2000, Vestergaard 2002) and the width of the
broad component of H$\beta$, summarized by: $M_{\rm BH}=4.9 \cdot 10^7
M_{\odot} L_{\rm 5100}^{0.5} W_{\rm 3000}^2$ (Shields et al.\ 2003),
where $L_{\rm 5100}$ is the luminosity of the continuum at 5100 \AA\,
in units of $10^{44}$ erg s$^{-1}$ and $W_{\rm 3000}$ is the FWHM of
the broad component H$\beta$ in units of 3000 \kms (Note that adopting
a different slope for $L_{\rm 5100}$, e.g. 0.66 in eq A5 in
Vestergaard 2002, does not change our result because of the luminosity
range spanned by our sample).  This formula assumes that the regions
emitting broad H$\beta$ in all AGN are photoionized under the same
conditions, by a UV continuum of the same shape. The bulk velocities
of the emitting gas clouds are presumed to be dominated by gravity, an
assumption for which there is currently moderate observational support
(Peterson \& Wandel 1999).

Absolute calibration of $L_{\rm 5100}$ was obtained by normalizing the
LRIS spectra to the model SDSS $r'$ magnitude.  Since SDSS photometry
and LRIS spectra are obtained at different epochs, the flux
normalization is uncertain by $\sim 20$ \% due to AGN variability
\cite{WM00}. This introduces a random error component of $\sim 10$\%
on \mbh, which is negligible with respect to the intrinsic scatter of
the ECPI method (a factor of $\sim 2.5$, Vestergaard 2002).  The
$L_{\rm 5100}$ flux was then corrected for Galactic extinction
(Schlegel et al.\ 1998) and subtracted the portion produced by
starlight, by comparing the measured line strength of the stellar
features with an assumed standard intrinsic value of
$\gamma=0.75\pm0.25$ (van der Marel 1994). The starlight falling
outside the slit, but included in the photometry, was estimated
from the measured seeing ($1\farcs0$) and a \dv\, profile for the host
galaxy with effective radius $0\farcs5$. The estimated fractions of
AGN-to-total light $f_{\rm AGN}$ are listed in Tab.~1.

The high signal-to-noise and resolution of the spectra allowed us to
determine the width of the broad component of H$\beta$ using the
following procedure (Fig.~2): {\it i)} the continuum was subtracted by
fitting a straight line between the continua at 4700 \AA\, and 5100
\AA\, rest frame. {\it ii)} The [\ion{O}{3}] line at 5007\AA\, was
divided by 3 and blueshifted to remove the 4959\AA\ line.  {\it iii)}
The [\ion{O}{3}] line at 5007\AA\, was blueshifted and rescaled to
remove the narrow component of H$\beta$. The line ratio H$\beta_{\rm
narrow}$ / [\ion{O}{3}]$\lambda$5007 was allowed to range between 1/20
and the maximum value consistent with the absence of ``dips'' in the
broad component (typically 1/10-1/7; e.g. Marziani et al.\ 2003). {\it
iv)} The second moment of the residual broad H$\beta$ component was
computed for the minimum and maximum narrow H$\beta$. The average and
semidifference of the two values were taken as best estimate and
uncertainty of the broad H$\beta$ rms. The rms was confirmed to be
much more insensitive than the FWHM to continuum and narrow component
subtraction (see also Peterson et al. 2004). {\it v)} The rms was
transformed into FWHM assuming FWHM = 2.35 rms, consistent with the
observed values (Peterson et al. 2004 report $2.03\pm0.59$. Their
value would lower \mbh\, by $\sim34$ \%).

\section{The Black Hole mass velocity dispersion relation at $ z\sim0.37$ }

The BHS relation is shown in Figure~\ref{fig:MBHs} and compared to the
local relations (Merritt \& Ferrarese 2001; Tremaine et al.\ 2002).
Local AGN with reliable $\sigma$ and \mbh\, from Ferrarese et
al. (2001) are also shown for comparison. We stress that the plot
should be interpreted with caution, since M$_{\rm BH}$ obtained with
the ECPI method is uncertain by a factor of $\sim 2.5$, as opposed to
the more precise estimates available for nearby objects. Furthermore,
the distribution of errors is non-gaussian and estimates for
individual objects may be significantly off, especially when based on
single epoch measurements (Vestergaard 2002). Given the relatively
large and non-gaussian uncertainties, selection effects could be
playing an important, and hard to quantify, role. Selection effects
enter in both quantities: $\sigma$ of bulges smaller or close to the
instrumental resolution ($\sigma_{s}$) cannot be reliably measured,
while very large $\sigma$ tend to dilute the absorption features and
therefore are harder to measure at a given S/N. Our observational
biases also limit the observable \mbh\, range. When we are
signal-to-noise limited, and therefore flux limited, we might exclude
objects with the smallest \mbh\, for a given $\sigma$. In contrast, by
requiring detectable stellar features in the integrated spectrum we
are excluding the brightest AGNs.

At face value, this result indicates that 4 Gyrs ago BHs lived in
bulges with lower $\sigma$.  The $\chi^2$ of the 7 points with respect
to the local relationship is 15.7, including errors on both
quantities, i.e. the probability that they are drawn from the local
relationship is 3\%.  Assuming that BH growth was negligible and
adopting the local BHS slope, the average difference and scatter in
the intercepts corresponds to $-0.16\pm0.13$ dex in $\log \sigma$ at
fixed \mbh.  Most of the offset and scatter is given by the galaxy
with the smallest $\sigma$ and largest fractional uncertainty
(S05). Discarding that point, the offset and scatter are
$-0.12\pm0.06$ dex in $\log\sigma$.  In terms of \mbh, the scatter
around the relation is 0.6 dex, or 0.3 dex excluding S05. Since the
estimated uncertainty on M$_{\rm BH}$ alone is $\sim0.4$ dex, the
intrinsic component of the scatter did not increase dramatically. Our
result appears to be at variance with that by Shields et al.\ (2003),
who combined ECPI \mbh\, with [\ion{O}{3}] emission line width, to
measure the BHS relation out to $z\sim3$. Their result is consistent
with no evolution, albeit with large scatter. More data are needed to
understand the origin and significance of this discrepancy. Possible
explanations include the different velocity dispersions adopted
(stellar features vs [\ion{O}{3}]) and the different range in \mbh. At
this stage we refrain from quantifying the slope.

In order to confirm this tantalizing and perhaps surprising result,
data for a larger and more complete sample will be needed, also to
address two important systematic effects that could conspire to
simulate the observed evolution. First, the local calibration of the
ECPI method might not be appropriate for the distant
universe. Reverberation mapping studies of distant galaxies could
verify this. Second, the assumed relationship between $\sigma$ and
$\sigma_{\rm ap}$ should be checked independently. For example the
contribution from a face-on cold disk could lower $\sigma_{\rm ap}$
requiring a higher value for ${\mathcal B}$, reducing the apparent
evolution: ${\mathcal B}$=1.59 would be required to bring the
$z\sim0.37$ and $z\sim0$ relationships in agreement. HST imaging is
needed to determine the inclination and relative luminosity of the
disk within the spectroscopic aperture, while higher signal-to-noise
ratio and spatially resolved spectroscopy are needed to study
departures from gaussianity of the line profile and rotational
support.

\section{Conclusions}

We have measured the velocity dispersion of the hosts of distant
active nuclei. In combination with \mbh\, estimates using the ECPI
method, we have obtained a first estimate of the BHS relation at
$z\sim0.37$. Since M$_{\rm BH}$ cannot decrease, this measurement,
taken at face value, suggests the $\sigma$ of bulges of Sy1s increased
by roughly 40\% in the past 4 Gyrs. Having provided a first
illustration of the method, it is now necessary to collect data for a
large sample of objects spanning a larger range of \mbh\, and
$\sigma$, and to test systematic uncertainties by obtaining
independent measures of \mbh\, and spatially resolved information on
the bulge and disk morphology and kinematics.

{\acknowledgments We thank A. Barth, G. Bertin, L.Ciotti, B.Hansen,
L.V.E. Koopmans, S. Gallagher, and the referee for useful
suggestions. The use of software developed by R.~P.~van der Marel is
gratefully acknowledged. TT is supported by the NASA Hubble Fellowship
grant HF-01167.01. This project relied upon the SDSS Database. The
authors acknowledge the role of Mauna Kea within the Hawaiian
community.}

\clearpage
\begin{deluxetable}{lccllllllll}
\tablecaption{Summary of relevant measurements \label{tab:sample}}
\tabletypesize{\tiny}
\tablehead{
\colhead{ID}  & \colhead{$\alpha_{2000}$} & \colhead{$\delta_{2000}$} & \colhead{S/N} & \colhead{$z$}    & \colhead{$\sigma_{\rm ap}$}        & \colhead{$\gamma$}       & \colhead{f$_{\rm 5100}$} & \colhead{rms(H$\beta$)}&  \colhead{f$_{\rm AGN}$} & \colhead{$\log$ (M$_{\rm BH}$/M$_{\odot}$)}}
\startdata
S01 & 15:39:16.23     & $+$03:23:22.06    & 36  & 0.3593 &         -       & 	 -	  &  6.00$\pm$0.01 & 50.0$\pm$0.4    & -    & - \\
S02 & 16:11:11.67     & $+$51:31:31.12    & 51  & 0.3543 &         -       & 	 -	  &  5.24$\pm$0.01 & 41.8$\pm$1.1    & -    & - \\
S03 & 17:32:03.11     & $+$61:17:51.95    & 65  & 0.3583 &         -       & 	 -	  & 10.72$\pm$0.02 & 35.7$\pm$0.2    & -    & - \\
S04 & 21:02:11.51     & $-$06:46:45.03    & 53  & 0.3577 &   140$\pm$17    &  0.34$\pm$0.13 &  7.77$\pm$0.01 & 50.2$\pm$1.1  & 0.44 & 8.18  \\
S05 & 21:04:51.84     & $-$07:12:09.45    & 54  & 0.3533 &    81$\pm$32    &  0.18$\pm$0.05 &  9.44$\pm$0.02 & 62.2$\pm$1.6  & 0.68 & 8.52  \\
S06 & 21:20:34.18     & $-$06:41:22.24    & 36  & 0.3686 &   123$\pm$31    &  0.24$\pm$0.09 &  8.85$\pm$0.02 & 45.6$\pm$0.6  & 0.58 & 8.21 \\
S07 & 23:09:46.14     & $+$00:00:48.91    & 65  & 0.3517 &         -       &       -        & 11.69$\pm$0.02 & 53.5$\pm$0.3  & -    & -  \\
S08 & 23:59:53.44     & $-$09:36:55.53    & 62  & 0.3583 &   139$\pm$23    &  0.21$\pm$0.08 &  8.62$\pm$0.01 & 27.2$\pm$0.5  & 0.63 & 7.77\\
S09 & 00:59:16.11     & $+$15:38:16.08    & 46  & 0.3539 &   137$\pm$30    &  0.22$\pm$0.08 & 10.80$\pm$0.02 & 41.5$\pm$1.1  & 0.62 & 8.17\\ 
S10 & 01:01:12.07     & $-$09:45:00.76    & 61  & 0.3509 &         -       &       -        & 14.46$\pm$0.02 & 42.1$\pm$0.8  & -    & - \\
S11 & 01:07:15.97     & $-$08:34:29.40    & 55  & 0.3555 &         -       &       -        &  9.07$\pm$0.02 & 32.8$\pm$0.1  & -    & - \\
S12 & 02:13:40.60     & $+$13:47:56.06    & 46  & 0.3579 &   161$\pm$36    &  0.26$\pm$0.07 & 13.70$\pm$0.03 & 72.0$\pm$0.7  & 0.55 & 8.66 \\
S99 & 16:00:02.80     & $+$41:30:27.00    & 49  & 0.3676 &   182$\pm$18    &  0.43$\pm$0.10 &  5.83$\pm$0.01 & 72.2$\pm$2.6  & 0.31 & 8.33 \\
\enddata

\tablecomments{For each object we list coordinates, average
signal-to-noise ratio per \AA\, of the region used for the kinematic
fit, redshift, velocity dispersion (in km s$^{-1}$), line strength,
total flux at 5100 \AA\, (rest frame; in 10$^{-17}$ erg
s$^{-1}$cm$^{-2}$\AA$^{-1}$; corrected as described in Section~3;
absolute flux calibration uncertainties are not included), rms width
of the broad component of H$\beta$ (in \AA, observed frame), estimated
fraction of AGN contribution to the flux at 5100\AA, estimated \mbh
(the related uncertainty is $\sim 0.4$ dex).}
\end{deluxetable}

\clearpage
\begin{figure}
\plotone{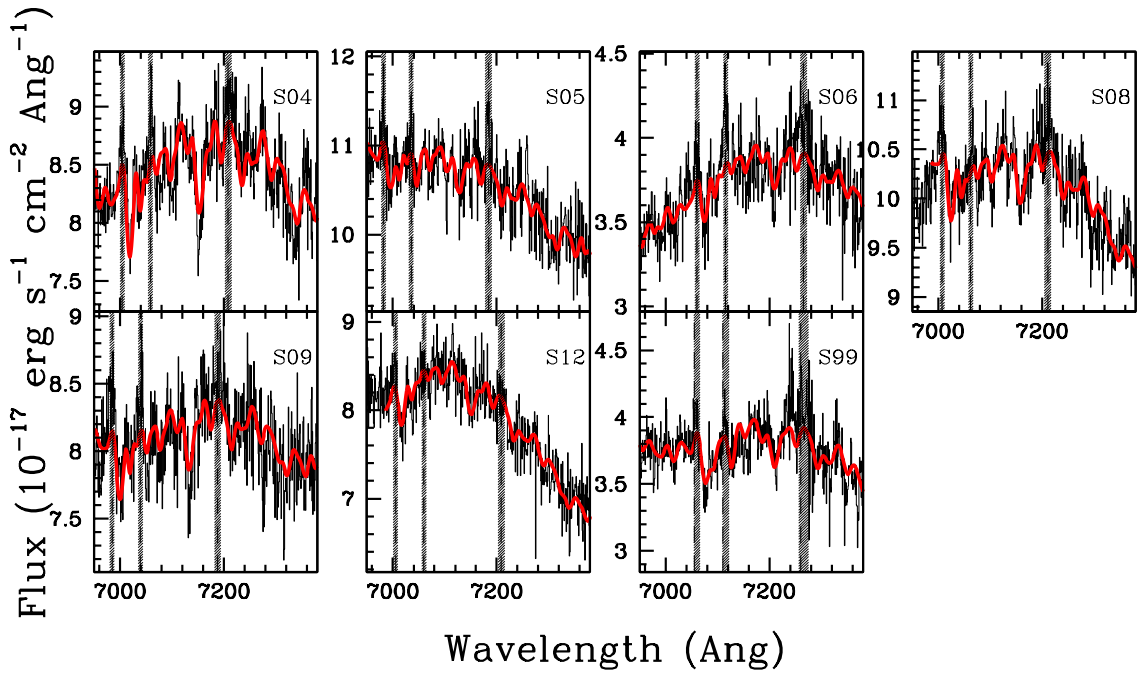}
\caption{Portion of the spectra used to measure bulge kinematics. The
black line is the data, the red lines is the best fit, shaded areas
indicate regions masked out during the fit.}
\label{fig:veldisp}
\end{figure}

\begin{figure}
\plotone{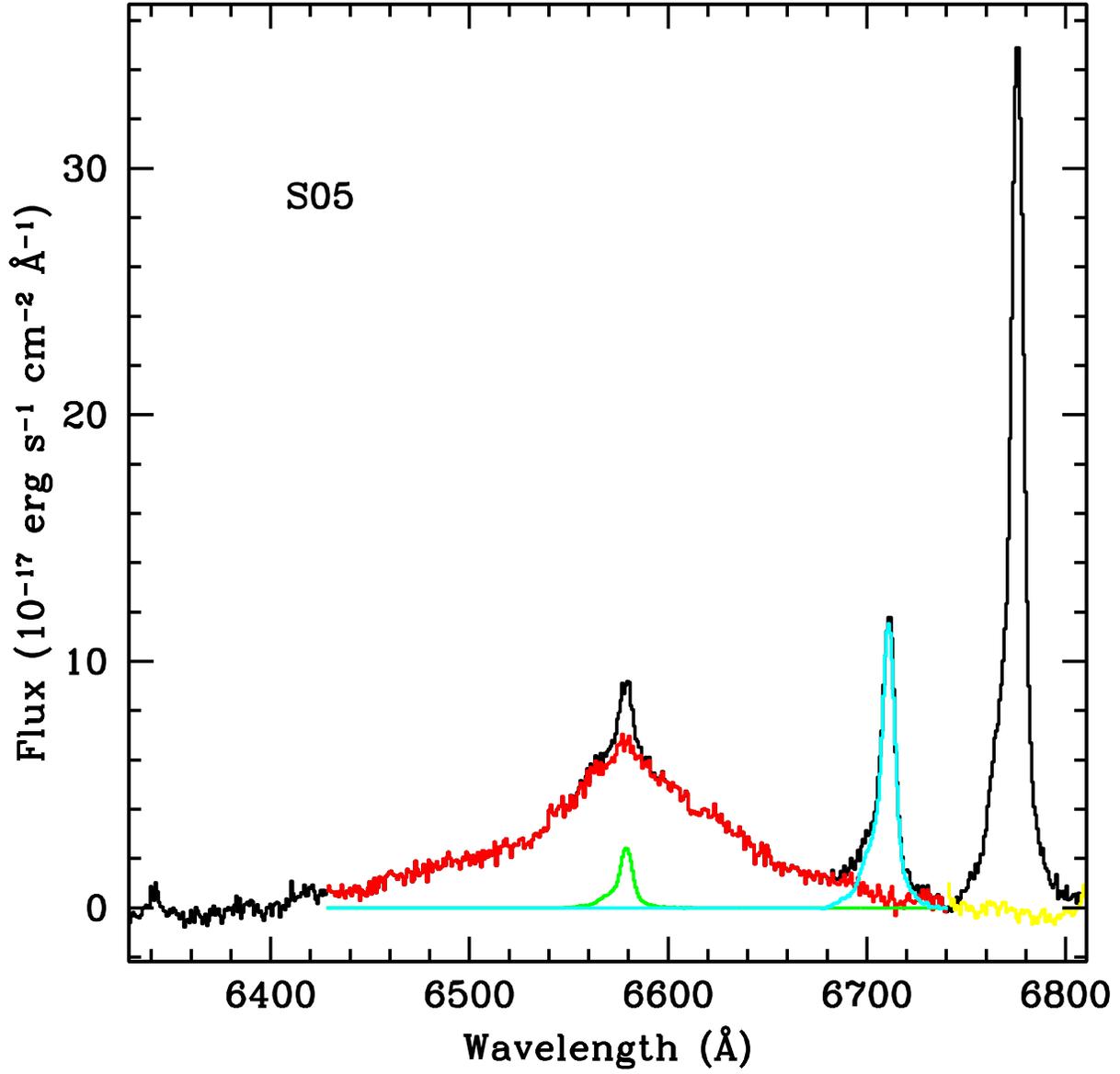}
\caption{Example of H$\beta$ width determination. The black line is
the original spectrum after continuum subtraction. The cyan and green
lines are the narrow components of [\ion{O}{3}]$\lambda$4959 and
H$\beta$ respectively, obtained by rescaling and blueshifting
[\ion{O}{3}]$\lambda$5007. The red spectrum is the residual broad line
used to compute the rms width. The yellow line underneath
[\ion{O}{3}]$\lambda$5007 is the reflection of the corresponding blue
part of H$\beta$ around its centroid.}
\label{fig:Hbeta}
\end{figure}

\begin{figure}
\plotone{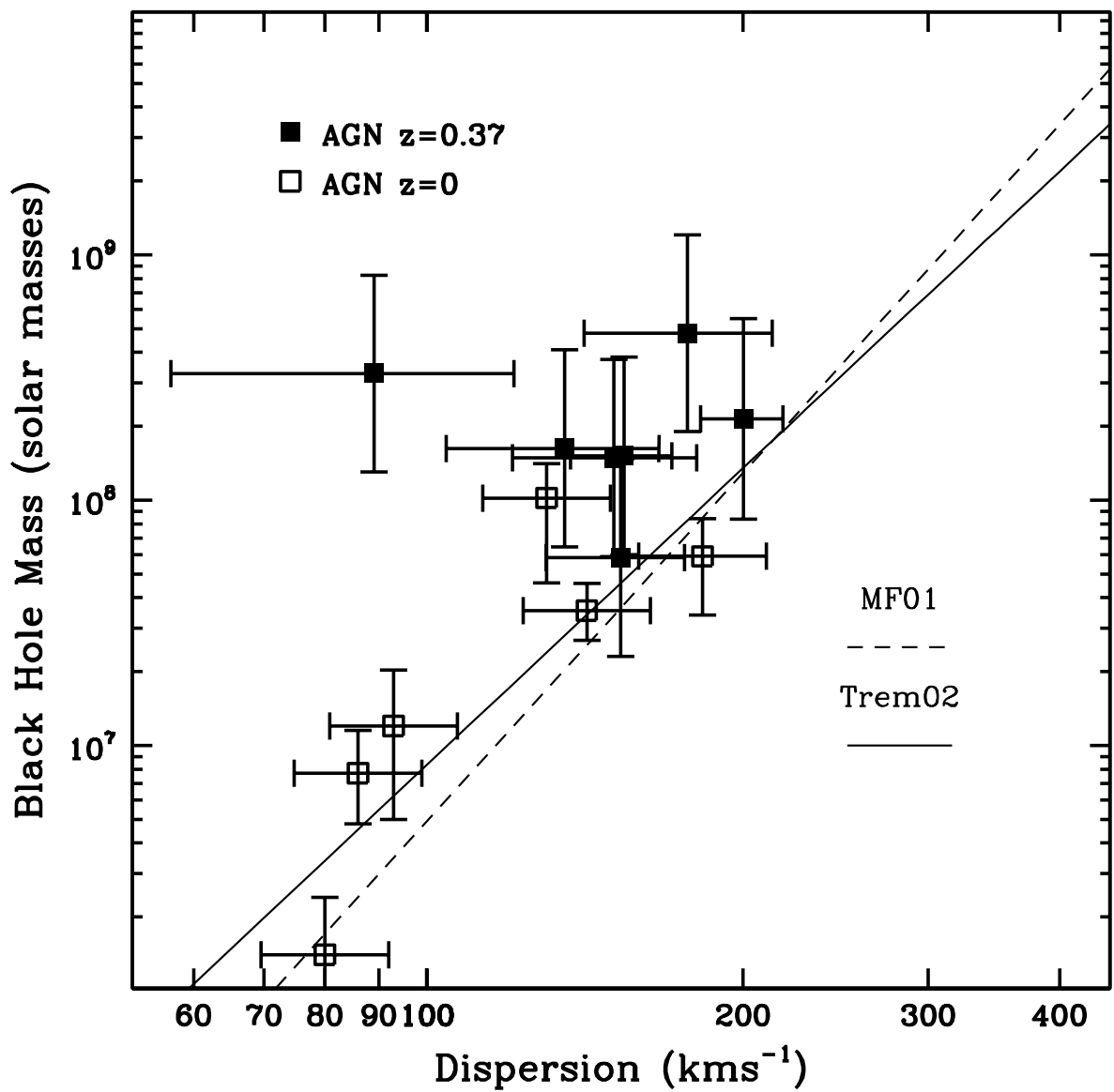}
\caption{Black hole mass velocity dispersion relation at
$z\sim0.37$ (solid squares with error bars). The local relations by
Merritt \& Ferrarese (2001) and Tremaine et al.\ (2002) are also shown
as solid and dashed lines. Since the latter adopts a slightly
different definition of velocity dispersion, it is overplotted
without corrections for comparison purposes only. Local AGN from
Ferrarese et al. (2001) are shown as open points.}
\label{fig:MBHs}
\end{figure}
\end{document}